# ARTICLE INFORMATION

**Article title**

Experimental Characterization Data for Battery Modules with Parallel-Connected Cells across Diverse Module-Level State of Health and Cell-to-Cell Variations

**Authors**

Qinan Zhou*, Daniel Stephens, Jing Sun*

**Affiliations**

Qinan Zhou is with the Department of Mechanical Engineering, University of Michigan, Ann Arbor, MI, 48103, USA.

Daniel Stephens and Jing Sun are with the Department of Naval Architecture and Marine Engineering, University of Michigan, Ann Arbor, MI 48103, USA.

**Corresponding author's email address and Twitter handle**

Qinan Zhou: qinan@umich.edu

Jing Sun: jingsun@umich.edu

**Abstract**

This experimental dataset presents both module-level and cell-level characterization data for lithium-ion battery modules composed of three parallel-connected inhomogeneous cells across a wide range of module-level state of health (M-SoH) and cell-to-cell variation (CtCV). First, 70 cells are aged to establish an inventory with cell-level state of health (C-SoH) ranging approximately from 100% to 80% (80% is considered as the end-of-life for automotive applications). From this inventory, 78 battery modules are then assembled, each exhibiting a distinct M-SoH value (from 100% to 80.98%) and a unique CtCV value (from 0% to 9.31%, defined as population standard deviation of C-SoH within each module). Module-level characterization data are collected under 25°C at 0.5C and 0.25C conditions, enabling extraction of module-level capacities and supporting diagnostic analyses such as incremental capacity analysis and differential voltage analysis. Before a module is assembled and tested, cell-level characterization tests are conducted for every individual cell within that module under 1C conditions, enabling direct quantification of CtCV and providing accurate labels for cell-level capacities and internal resistances. The dataset is organized with both raw time-series data and processed summary information such as C-SoH, M-SoH, and CtCV for all modules. With the paired module-level and cell-level characterization data, this dataset enables understanding and development of advanced degradation monitoring mechanisms for battery modules with parallel-connected cells in the presence of CtCVs.

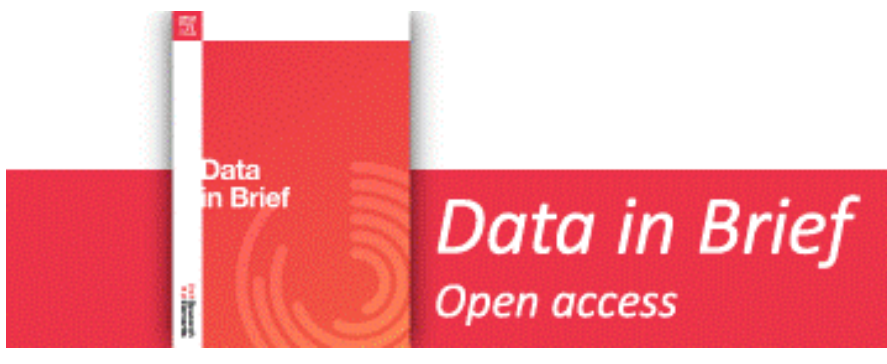

## SPECIFICATIONS TABLE

| | |
|---|---|
| **Subject** | Engineering & Materials science |
| **Specific subject area** | Module and Cell Characterization Data for Battery Modules with Parallel-Connected Cells Cross a Wide Span of State of Health and Cell-to-Cell Variations |
| **Type of data** | Raw and Processed Data |
| **Data collection** | **(1) Module Characterization Tests:**<br>Tests are performed for 78 modules. All modules consist of 3 parallel-connected inhomogeneous cells, exhibiting a wide span of module-level state of health (M-SoH) and cell-to-cell variation (CtCV). Maccor S4000 executes constant-current-constant-voltage (CC-CV) charging and constant-current (CC) discharging at 0.5C and 0.25C under 25°C ambient temperature.<br>**(2) Cell Characterization Tests:**<br>Tests are conducted on every cell within each of 78 modules, providing direct quantification of CtCV within battery modules. Maccor S4000 executes CC-CV charging, CC discharging, and pulsed charging/discharging tests at 1C and 25°C ambient temperature. |
| **Data source location** | U-M Battery Lab, University of Michigan, Ann Arbor, Michigan, United States of America |
| **Data accessibility** | Repository name: Mendeley Data<br>Data identification number: 10.17632/ssrgfmb8vw.2<br>Direct URL to data: https://data.mendeley.com/datasets/ssrgfmb8vw<br>When using the data, users should cite this dataset descriptor, rather than the dataset itself on Mendeley Data. |
| **Related research article** | None |

## VALUE OF THE DATA

- This dataset provides characterization data for both battery modules and their constituent cells across a wide range of module-level state of health (M-SoH) and cell-to-cell variation (CtCV). Each module consists of three parallel-connected inhomogeneous cells.
- The module characterization data includes 78 modules, each exhibiting unique M-SoH and CtCV. Characterizations are conducted under 25°C ambient temperature at 0.25C and 0.5C conditions, enabling extraction of key information such as module-level capacity and supporting diagnostic analyses including incremental capacity analysis and differential voltage analysis.
- Cell characterization data are provided for every individual cell within each module, enabling direct quantification of CtCV and providing labels for cell-level capacities and internal resistances within all 78 modules. Cell characterizations are conducted under 25 °C ambient temperature at 1C conditions.
- To the best of our knowledge, this is the first large experimental dataset that includes both module-level and cell-level characterization data, supporting battery module-level degradation research.

## BACKGROUND

Practical lithium-ion battery systems require individual cells to be connected in series and/or parallel to form modules and packs that meet energy and power requirements [1]. This dataset focuses on battery modules with parallel-connected inhomogeneous cells. Module-level degradation monitoring involves two key concepts: module-level state of health (M-SoH) and cell-to-cell variation (CtCV). While many datasets exist for cell-level degradation [2], to the best of our knowledge, no experimental datasets are available in public domain that have systematically captured both M-SoH and CtCV across a diverse population of battery modules, hindering research in module degradation monitoring. To address this need, this work presents a large experimental dataset specifically designed for module-level degradation analysis. It includes characterization data for 78 modules, spanning a wide range of M-SoH and CtCV conditions, at two levels of C-rates under 25°C ambient temperature. Crucially, it also provides detailed characterization data for every individual cell within each module, enabling direct quantification of CtCV and offering labels for cell-level capacities and internal resistances within all 78 modules. This dataset has been used by [3] to develop algorithms for estimation of M-SoH and CtCV using only module-level measurements.

## DATA DESCRIPTION

This dataset comprises characterization data for 78 battery modules with a diverse range of M-SoH and CtCV values, collected under an ambient temperature of 25°C at two C-rates: 0.5C and 0.25C. Each module consists of 3 parallel-connected inhomogeneous cells, whose cell-level state of health (C-SoH) values range approximately from 100% to 80%. Note that the lower bound of 80% is chosen as it is a consented indicator of the end-of-life for automotive applications. Furthermore, this dataset provides separate characterization data for every individual cell within all modules, providing information on cell-level capacities and internal resistances for direct quantification of CtCV within battery modules.

This section first defines the quantification metrics for cell-level state of health (C-SoH), module-level state of health (M-SoH), and cell-to-cell variation (CtCV) used in this dataset. It then describes the battery cells and modules, followed by a detailed account of the characterization data provided for both modules and their constituent cells. Finally, it explains the folder structures of the dataset.

### A. Definitions

For SoH metrics, this dataset focuses on capacity fading. Following the standard definition [1], this dataset defines the C-SoH as:

$$\mathrm{SoH}_c = \frac{C_c}{C_{c,\mathrm{fresh}}}, \tag{1}$$

where subscript $c$ denotes cell-level quantities, $\mathrm{SOH}_c$ is the C-SoH, $C_c$ and $C_{c,\mathrm{fresh}}$ are the current and fresh cell-level capacities, respectively. This dataset focuses on battery modules with parallel-connected cells, and the M-SoH is defined as:

$$\mathrm{SoH}_m = \frac{C_m}{C_{m,\mathrm{fresh}}}, \tag{2}$$

where subscript $m$ denotes module-level quantities, $\mathrm{SOH}_m$ is the M-SoH, $C_m$ and $C_{m,\mathrm{fresh}}$ are the current and fresh module-level capacities, respectively. Setting the fresh cell-level capacity equal to the nominal capacity for all cells within a module, M-SoH defined in Equation (2) can also be expressed as:

$$\mathrm{SoH}_m = \frac{\sum_{i=1}^{N_p} C_{c,i}}{N_p C_{c,\mathrm{fresh}}} = \mathrm{mean}\left(\{\mathrm{SoH}_{c,i}\}\right), \tag{3}$$

where $N_p$ is the number of parallel-connected cells, $C_{c,i}$ denotes the cell-level capacity of the $i$-th cell inside the module, and $\left\{\mathrm{SoH}_{c,i}\right\} = \left\{\mathrm{SoH}_{c,1}, \mathrm{SoH}_{c,2}, \dots, \mathrm{SoH}_{c,N_p}\right\}$. For this dataset, $N_p = 3$.

While the standard metrics have been used for SoH, there is no universally accepted metric for quantifying CtCVs. Supporting the work of [3], this dataset focuses on the CtCV in C-SoH values and adopts the population standard deviation (SD) as the metric, defined as:

$$\mathrm{SD} = \mathrm{sd}\left(\{\mathrm{SoH}_{c,i}\}\right). \tag{4}$$

Subsection B will provide specific examples to build intuitions about the scale of SD as a CtCV metric and how SD related to the underlying C-SoH values within a battery module.

## B. Battery Modules and Cells

This dataset is constructed from an inventory of 70 cylindrical Sony US18650VTC6 cells with C-SoH values ranging approximately from 100% to 80%. Their aging procedures are described in the *Experimental Design, Materials, and Methods* Section. While the complete datasheet of the cell is provided in [4], key specifications are summarized in Table 1. Using scanning electron microscopy and energy-dispersive X-ray spectroscopy, prior studies have identified the positive electrode as lithium nickel-cobalt-aluminum oxide (NCA) and the negative electrode as silicon-graphite [5].

**Table 1. Key Specifications of Battery Cells**

| | |
|---|---|
| **No. of Cells** | 70 |
| **Positive Electrode Chemistry** | NCA [5] |
| **Negative Electrode Chemistry** | Graphite + Si [5] |
| **Shape, Size** | Cylindrical, 18650 |
| **Nominal Capacity** | 3Ah |
| **Span of Cell-Level SoH** | Approximately, 100% - 80% |
| **Charge Cutoff Voltage** | 4.2V |
| **Discharge Cutoff Voltage** | 2.5V |

This dataset comprises 78 battery modules spanning a wide range of M-SoH and CtCV values. Each module is assembled by connecting three of these cells in parallel, as illustrated in Fig. 1. Each cell within a module has a distinct C-SoH value, introducing CtCVs into the module. Table 2 summarizes key specifications of these modules, while Fig. 2 presents the distribution of M-SoH and CtCV values across all 78 modules. These modules are constructed to intentionally span a wide range of M-SoH and CtCV values, enabling a comprehensive investigation of module behavior under diverse conditions. This broad coverage also facilitates the development, validation, and evaluation of different battery module management algorithms across a variety of scenarios.

**Table 2. Key Specifications of Battery Modules**

| | |
|---|---|
| **No. of Modules** | 78 |
| **Module Configuration** | 3 Cells in Parallel |
| **Nominal Capacity** | 9Ah |
| **Span of Module-Level SoH** | 100% - 80.98% |
| **Span of Cell-to-Cell Variation** | 0% - 9.31% SD |
| **Charge Cutoff Voltage** | 4.2V |
| **Discharge Cutoff Voltage** | 2.5V |

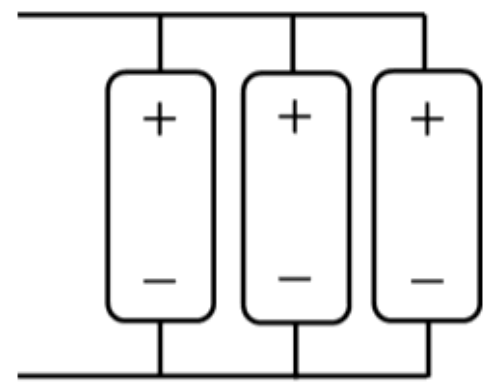

**Fig. 1. Battery Module Configuration**

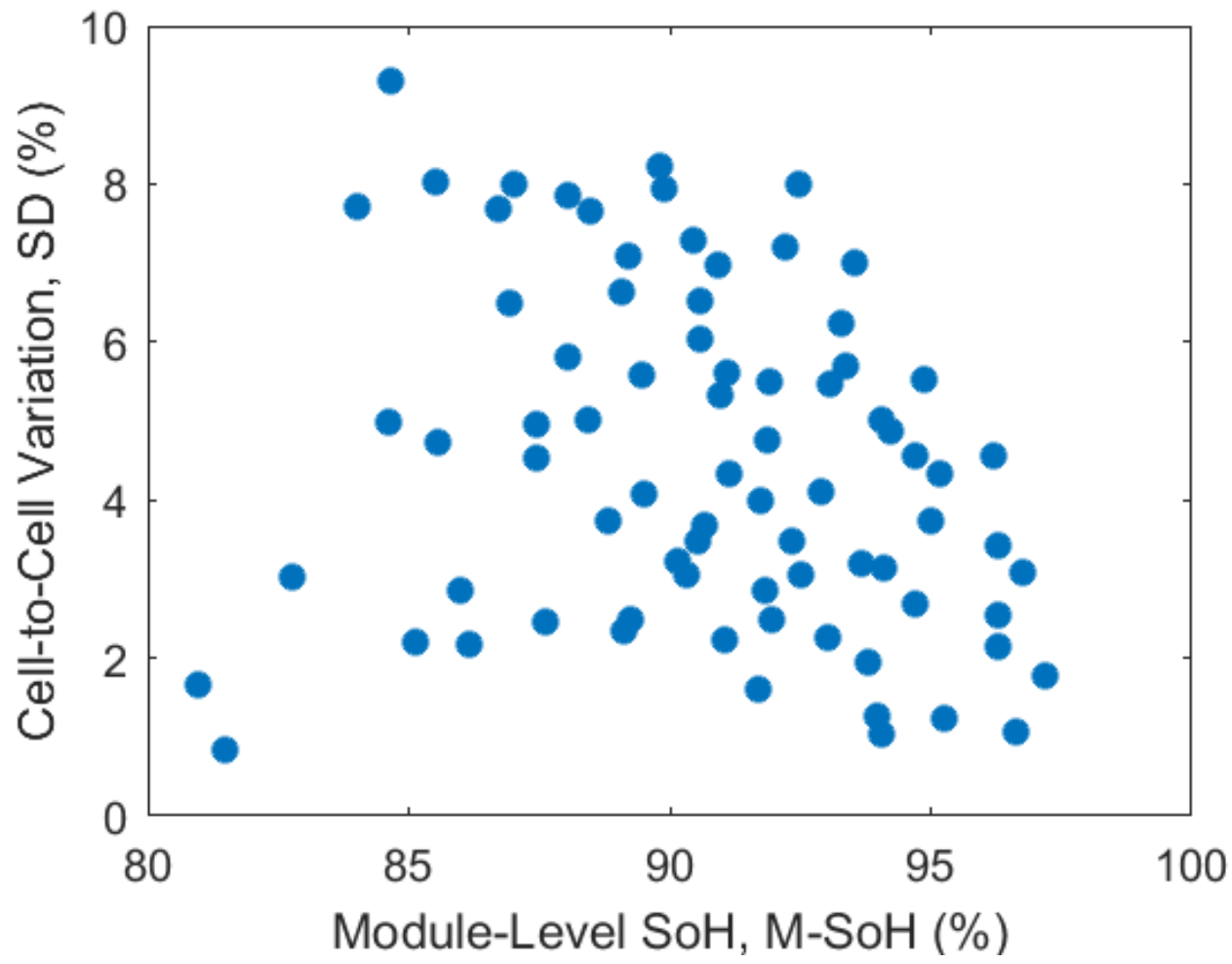

**Fig. 2. Distribution of Cell-to-Cell Variations and Module-Level SoH Values inside Dataset**

Table 3 summarizes CtCV, M-SoH, and C-SoH values for several representative modules in the dataset. Based on Table 3, the module with the maximum CtCV in the dataset exhibits a difference of approximately 21% in the C-SoH values, while the module with the median CtCV still shows a difference of about 10% in C-SoH values. Thus, the dataset encompasses modules with a broad range of CtCV levels. Table 3 also helps build intuition regarding the scale of SD as a CtCV metric.

**Table 3. Example Modules with Corresponding Module-Level SoH and Cell-to-Cell Variations**

| Cell-to-Cell Variation | Module-Level SoH | Cell-Level SoH |
|---|---|---|
| 9.31% SD<br>(Maximum CtCV in Dataset) | 84.65% SoH | 99.2% SoH, 80.6% SoH, 78.5% SoH |
| 4.54% SD<br>(Median CtCV in Dataset) | 88.67% SoH | 91.9% SoH, 91.8% SoH, 82.3% SoH |
| 0.84% SD<br>(Minimum CtCV in Dataset) | 81.46% SoH | 83.0% SoH, 81.9% SoH, 80.9% SoH |

**Table 4. Complete Information on Module-Level SoH and Cell-to-Cell Variations for All the Modules**

| No. | M-SoH (%) | CtCV (%) | C-SoH (%) | No. | M-SoH (%) | CtCV (%) | C-SoH (%) | No. | M-SoH (%) | CtCV (%) | C-SoH (%) |
|---|---|---|---|---|---|---|---|---|---|---|---|
| 01 | 89.07 | 6.63 | 81.3, 94.6, 96.1 | 27 | 86.00 | 2.85 | 83.6, 87.5, 90.6 | 53 | 90.94 | 6.97 | 83.0, 94.8, 99.6 |
| 02 | 90.16 | 3.23 | 86.7, 92.2, 94.4 | 28 | 80.98 | 1.67 | 80.2, 81.2, 84.1 | 54 | 93.54 | 7.01 | 85.0, 98.8, 100.8 |
| 03 | 96.30 | 3.42 | 92.9, 99.2, 100.8 | 29 | 95.27 | 1.23 | 95.0, 96.9, 98.0 | 55 | 86.91 | 6.48 | 80.2, 88.0, 96.0 |
| 04 | 95.02 | 3.74 | 91.0, 97.9, 99.7 | 30 | 94.05 | 1.04 | 93.8, 95.9, 96.1 | 56 | 84.03 | 7.71 | 78.8, 81.1, 96.2 |
| 05 | 94.12 | 3.14 | 91.0, 97.4, 97.9 | 31 | 93.95 | 1.25 | 93.7, 94.6, 96.7 | 57 | 90.57 | 6.51 | 82.9, 94.1, 98.3 |
| 06 | 92.34 | 3.49 | 88.7, 94.3, 97.1 | 32 | 91.69 | 1.60 | 90.7, 93.5, 94.4 | 58 | 93.26 | 6.25 | 85.7, 97.9, 99.8 |
| 07 | 90.52 | 3.47 | 86.5, 92.9, 94.6 | 33 | 96.63 | 1.06 | 97.4, 97.8, 99.8 | 59 | 90.57 | 6.03 | 83.3, 94.5, 97.2 |
| 08 | 88.06 | 5.81 | 81.6, 91.6, 95.4 | 34 | 89.25 | 2.48 | 92.2, 91.9, 86.8 | 60 | 91.09 | 5.60 | 84.8, 93.4, 98.3 |
| 09 | 94.71 | 2.70 | 92.3, 96.6, 98.8 | 35 | 81.46 | 0.84 | 80.9, 81.9, 83.0 | 61 | 89.45 | 5.57 | 83.3, 91.0, 96.9 |
| 10 | 93.69 | 3.19 | 91.0, 94.7, 98.8 | 36 | 88.43 | 5.01 | 83.3, 91.8, 95.2 | 62 | 93.37 | 5.68 | 87.0, 96.9, 100.4 |
| 11 | 92.52 | 3.06 | 90.0, 93.4, 97.5 | 37 | 86.16 | 2.18 | 84.4, 87.6, 89.7 | 63 | 93.05 | 5.47 | 86.7, 96.4, 99.6 |
| 12 | 91.82 | 2.85 | 89.0, 93.4, 95.9 | 38 | 84.62 | 4.99 | 78.8, 87.5, 90.5 | 64 | 91.89 | 5.48 | 85.7, 95.2, 98.7 |
| 13 | 90.34 | 3.06 | 87.7, 91.9, 95.2 | 39 | 85.57 | 4.73 | 81.6, 85.5, 93.0 | 65 | 90.97 | 5.32 | 84.8, 94.3, 97.2 |
| 14 | 96.74 | 3.08 | 94.1, 100.5, 100.6 | 40 | 82.78 | 3.03 | 80.2, 83.3, 87.6 | 66 | 94.86 | 5.52 | 89.0, 100.7, 100.7 |
| 15 | 87.43 | 4.97 | 82.1, 90.5, 94.0 | 41 | 85.51 | 8.02 | 80.4, 81.2, 97.8 | 67 | 94.06 | 5.02 | 88.6, 98.7, 99.8 |
| 16 | 91.95 | 2.48 | 87.0, 90.1, 93.0 | 42 | 89.82 | 8.22 | 79.9, 93.9, 99.5 | 68 | 94.24 | 4.88 | 88.7, 97.9, 99.9 |
| 17 | 96.29 | 2.16 | 94.6, 98.6, 99.5 | 43 | 87.02 | 8.00 | 79.9, 86.0, 99.1 | 69 | 91.85 | 4.76 | 86.5, 95.4, 97.5 |
| 18 | 96.28 | 2.55 | 94.0, 98.8, 99.9 | 44 | 92.46 | 8.00 | 82.3, 98.7, 99.7 | 70 | 91.13 | 4.34 | 86.6, 92.9, 97.2 |
| 19 | 93.78 | 1.96 | 92.3, 96.2, 96.7 | 45 | 89.90 | 7.95 | 80.9, 95.2, 99.5 | 71 | 95.18 | 4.33 | 90.6, 97.9, 100.9 |
| 20 | 93.04 | 2.26 | 91.3, 94.2, 96.8 | 46 | 86.73 | 7.68 | 80.2, 84.9, 98.3 | 72 | 94.68 | 4.55 | 89.9, 96.7, 100.9 |
| 21 | 91.06 | 2.22 | 89.0, 93.5, 93.9 | 47 | 84.65 | 9.31 | 80.6, 78.5, 99.2 | 73 | 92.91 | 4.12 | 88.9, 94.7, 98.9 |
| 22 | 87.44 | 4.54 | 82.3, 91.8, 91.9 | 48 | 88.03 | 7.86 | 80.4, 87.7, 99.5 | 74 | 91.73 | 4.00 | 87.6, 93.6, 97.3 |
| 23 | 87.63 | 2.45 | 85.0, 89.6, 90.7 | 49 | 88.49 | 7.66 | 79.0, 94.2, 96.2 | 75 | 89.52 | 4.07 | 84.9, 91.9, 94.5 |
| 24 | 85.11 | 2.19 | 82.9, 87.0, 88.0 | 50 | 90.45 | 7.28 | 82.0, 93.1, 99.6 | 76 | 96.20 | 4.55 | 91.0, 100.7, 100.7 |
| 25 | 89.12 | 2.35 | 88.0, 89.6, 93.6 | 51 | 89.19 | 7.08 | 81.0, 91.5, 98.2 | 77 | 88.80 | 3.73 | 85.0, 90.7, 94.1 |
| 26 | 97.19 | 1.77 | 96.7, 98.0, 100.9 | 52 | 92.20 | 7.19 | 83.3, 96.1, 100.1 | 78 | 90.68 | 3.69 | 86.8, 92.9, 95.6 |

Complete information on the CtCV, M-SoH, and C-SoH values for all 78 modules are listed in Table 4. For simplicity, the C-SoH and M-SoH values are calculated by setting the fresh capacities in Equations (1) and (2) equal to the nominal capacities listed in Tables 1 and 2, respectively. Because some cells may have slightly higher fresh capacities than the nominal value due to manufacturing variations, the computed C-SoH can occasionally exceed 100% under this convention.

*Remark 1.* Constructing 78 modules from a pool of 70 cells inevitably requires the re-use of certain cells. As to be discussed in Subsection B.2 of *Experimental Design, Materials, and Methods* Section, prior to re-using, cells follow a designed re-characterization schedule that ensures the unobserved C-SoH decrease in re-used cells is at most around 1%. Nonetheless, this may introduce an inherent limitation to the dataset.

### C. Characterization Data of Battery Modules and Cells

This dataset provides characterization data for all 78 battery modules spanning a wide range of M-SoH and CtCV values, collected under 25°C ambient temperature at two C-rates: 0.5C and 0.25C. The sampling frequency is 1Hz. Cell and module temperatures are not actively monitored during testing. However, because all characterization tests were performed at relatively low current rates (0.25C and 0.5C), and the laboratory is equipped with an effective ventilation system, the resulting temperature rise is expected to be limited. For the sign convention, all the tests treat the charging current as positive.

Fig. 3 illustrates the characterization data for a representative module. As shown, each module undergoes three cycles. The first cycle serves to standardize initial conditions, ensuring a common baseline across modules with different initial states. The second and third cycles correspond to characterization tests conducted at 0.5C and 0.25C, respectively. These two cycles enable extraction of key information, such as M-SoH, and support diagnostic analyses, including incremental capacity analysis and differential voltage analysis. Detailed experimental procedures are provided in the *Experimental Design, Materials, and Methods* Section.

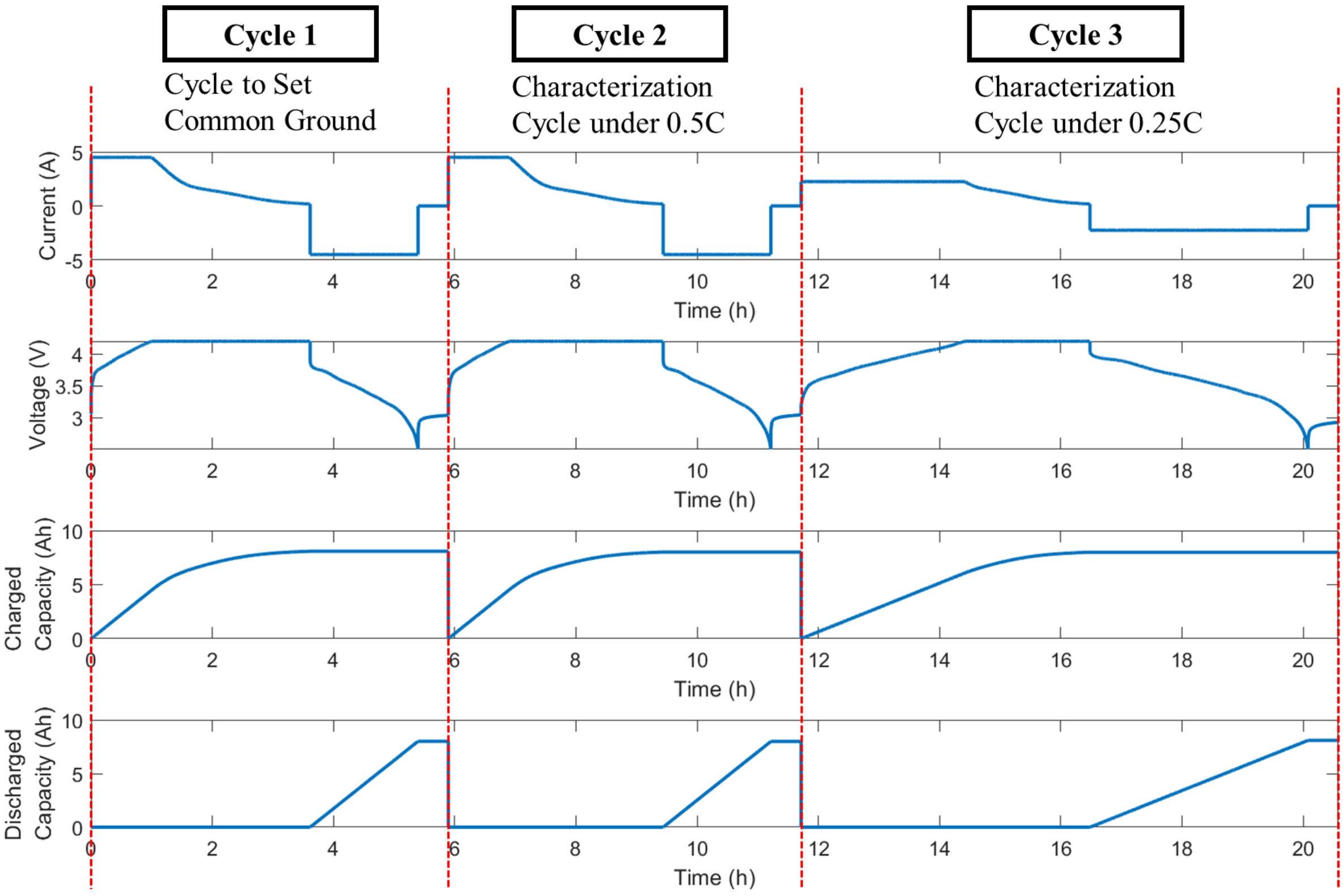

**Fig. 3. Example Module Characterization Data at 0.5C and 0.25C under 25°C Ambient Temperature**

This dataset also provides separate characterization data for all three individual cells within each module, collected at 25°C ambient temperature under 1C. Similarly, the sampling frequency is 1Hz. The cell temperatures are not monitored, and the resulting temperature rise is expected to be limited. Fig. 4 shows the characterization data for a representative cell. As shown, each cell undergoes four main steps. The first step standardizes initial conditions, establishing a common baseline across cells with different initial states. The second and third steps correspond to characterization under 1C discharging and charging, respectively. These tests enable extraction of key metrics, for example C-SoH. The fourth step provides an approximate estimation of cell-level internal resistance. Due to practical constraints, this step is simplified from a standard hybrid pulse power characterization (HPPC) test. Experimental procedures and associated limitations of cell characterization data are detailed in the *Experimental Design, Materials, and Methods* Section.

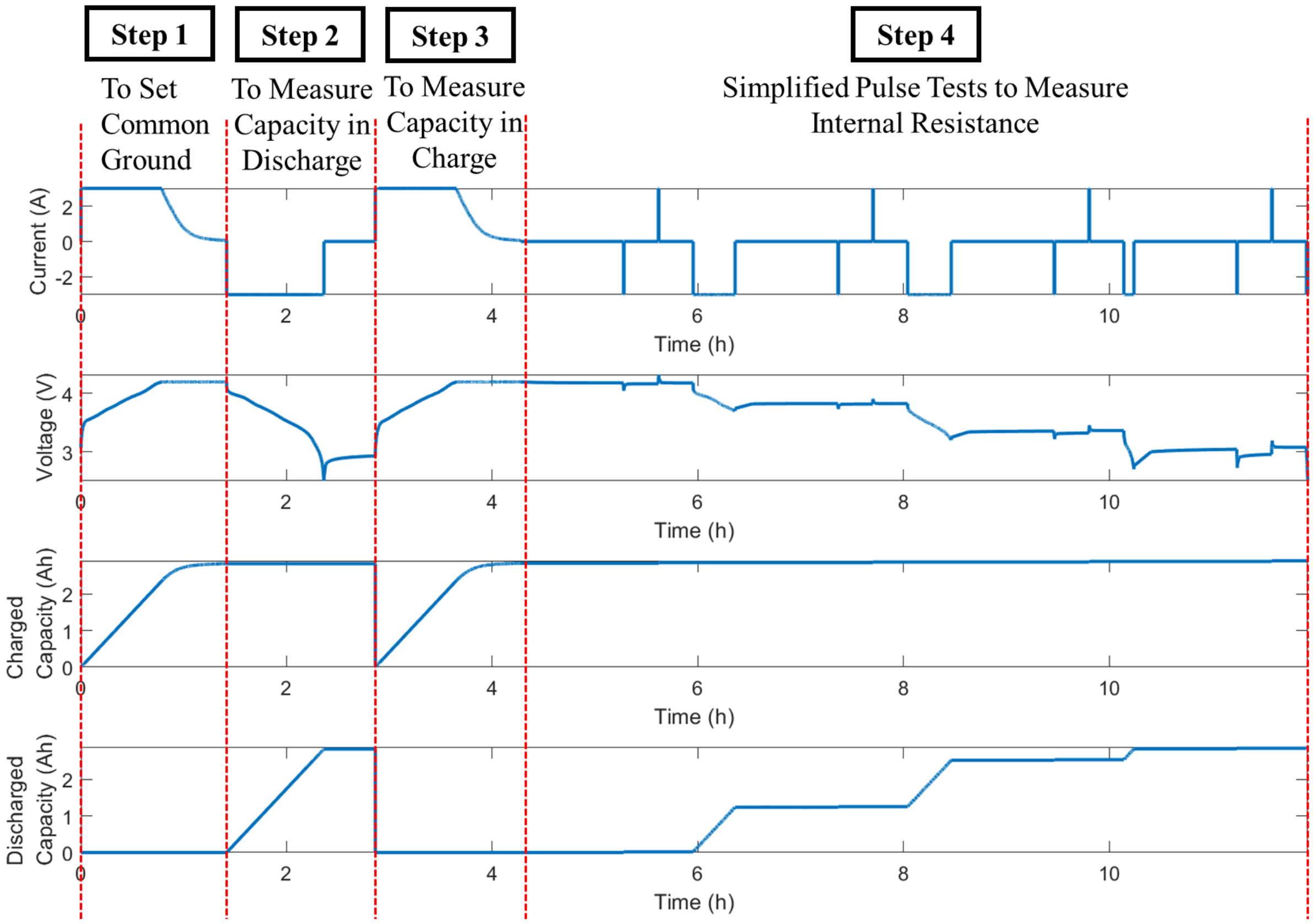

**Fig. 4. Example Cell Characterization Data at 1C under 25°C Ambient Temperature**

### D. Folder Structure of the Dataset

Fig. 5 illustrates the folder structure of the dataset. The top-level directory, “Dataset,” contains a “Readme.pdf” file, a “license.txt” file, a “original_hashes.csv” file, a “copy_hashes.csv” file, a “citation_instruction” file, and a “Characterization” folder. The “Readme.pdf” file provides detailed documentation of the dataset, while the “Characterization” folder contains all characterization data for both battery modules and their individual cells, as described in Subsections B and C. The "license.txt" file contains the dataset license statement. The "original_hashes.csv" and "copy_hashes.csv" files contain the SHA-256 hashes of all data files in the author's original copy and the copy downloaded from the website repository, respectively. A comparison of these hashes confirms that the files are identical, demonstrating that the integrity of the dataset has been preserved during distribution. Finally, the “citation_instruction.txt” file provides guidance on how the dataset should be cited in future publications.

Within the “Characterization” folder, there are 78 subfolders, each corresponding to one module and its associated three cells. The naming convention follows “Module_x_and_its_Cells,” where x denotes the module index. Each of these subfolders contains two directories: “Raw” and “Summary.” The “Raw” folder stores the experimental data from characterization tests, while the “Summary” folder provides summary information.

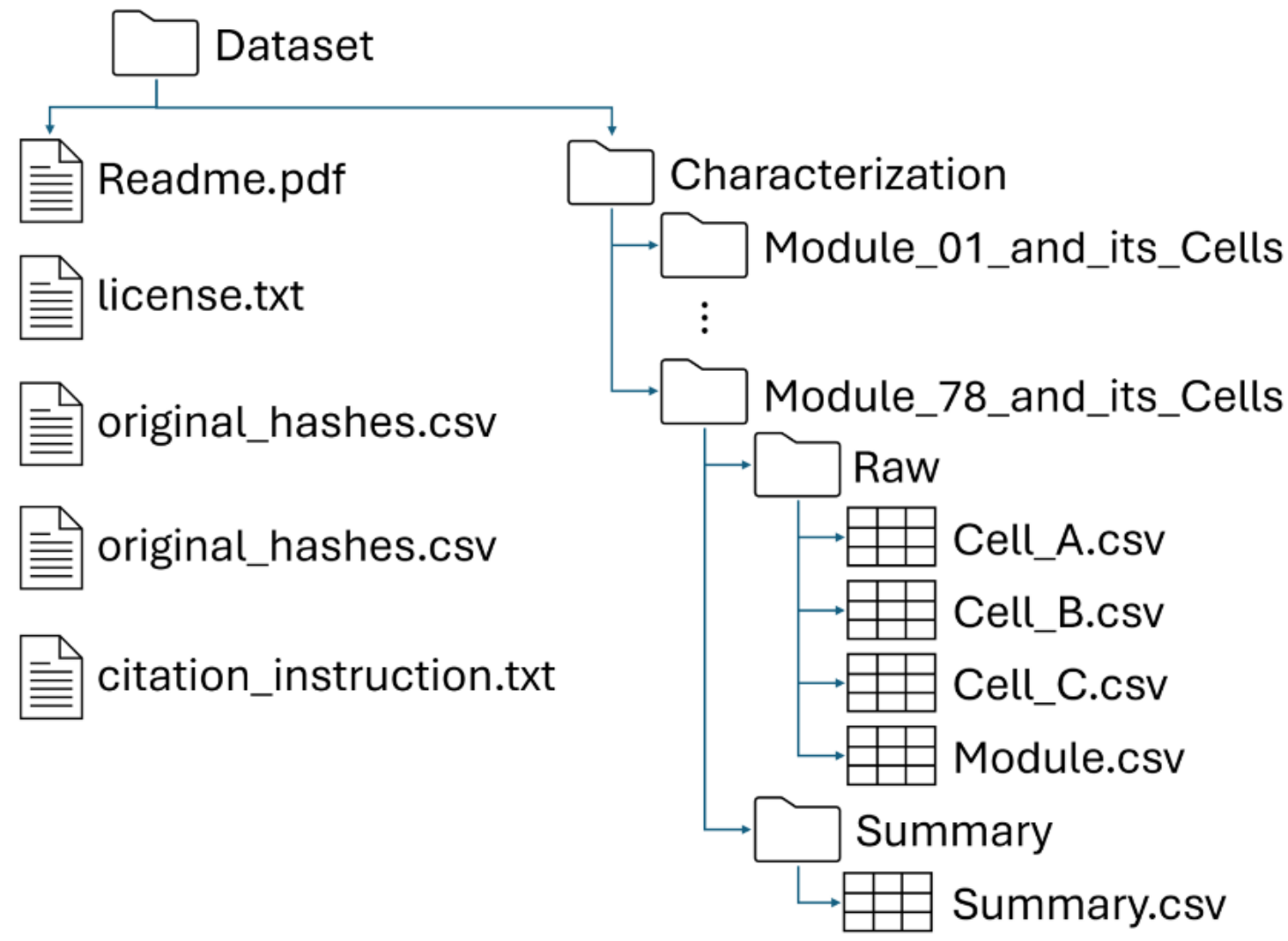

**Fig. 5. Folder Structure of the Dataset**

Inside the "Raw" folder, the files "Cell_A.csv," "Cell_B.csv," and "Cell_C.csv" contain characterization data for the three cells within a module (as shown in Fig. 4), and "Module.csv" contains the corresponding module-level characterization data (as shown in Fig. 3). These four files share the same headers. The headers of these files, along with their physical meanings and units, are summarized in Table 5.

**Table 5. Header Information for Cell_A.csv, Cell_B.csv, Cell_C.csv, and Module.csv Files**

| Header Name | Physical Meaning | Units |
|---|---|---|
| Current_A | Current | A |
| Voltage_V | Voltage | V |
| Time_s | Time | Second |
| Charged_Capacity_Ah | Charged Capacity | Ah |
| Discharged_Capacity_Ah | Discharged Capacity | Ah |
| Cycle_Index | Index for Cycles | unitless |

Within the "Summary" folder, the file "Summary.csv" provides aggregated information for the module and its three cells, including C-SoH, M-SoH, and CtCV values. The headers in this file, along with their physical meanings and units, are summarized in Table 6.

**Table 6. Header Information for Summary.csv File**

| Header Name | Physical Meaning | Units |
|---|---|---|
| Cell_A_Capacity_Ah | Capacity of Cell A | Ah |

| Cell_A_SOH_Percent | SoH of Cell A | % |
|---|---|---|
| Cell_B_Capacity_Ah | Capacity of Cell B | Ah |
| Cell_B_SOH_Percent | SoH of Cell B | % |
| Cell_C_Capacity_Ah | Capacity of Cell C | Ah |
| Cell_C_SOH_Percent | SoH of Cell C | % |
| Module_Capacity_Ah | Capacity of Module | Ah |
| Module_SOH_Percent | SoH of Module | % |
| Population_Standard_Deviation_of_Cell_SOH_Percent | Cell-to-Cell Variation | % |

# EXPERIMENTAL DESIGN, MATERIALS AND METHODS

Fig. 6 provides a high-level flowchart of experimental procedures to create this dataset. Based on Fig. 6, the process begins with aging cycles applied to 70 fresh cells purchased from a manufacturer to generate an inventory with diverse C-SoH values. These cells are then characterized to measure their capacities and internal resistances under 25 °C ambient temperature. Next, three cells with different C-SoH values are assembled into a battery module, and module-level characterization tests are conducted at 0.5C and 0.25C under 25°C ambient temperature. Finally, the tested module is disassembled. This procedure is repeated for 78 modules, each exhibiting distinct M-SoH and CtCV conditions. To account for ongoing cell degradation during repeated testing and accommodate for the budget, periodic cell re-characterization is performed according to a predefined empirical schedule. Each step in Fig. 6, along with the predefined empirical schedule for cell re-characterization, is detailed in the following subsections. All the experimental data are recorded at a sampling frequency of 1Hz.

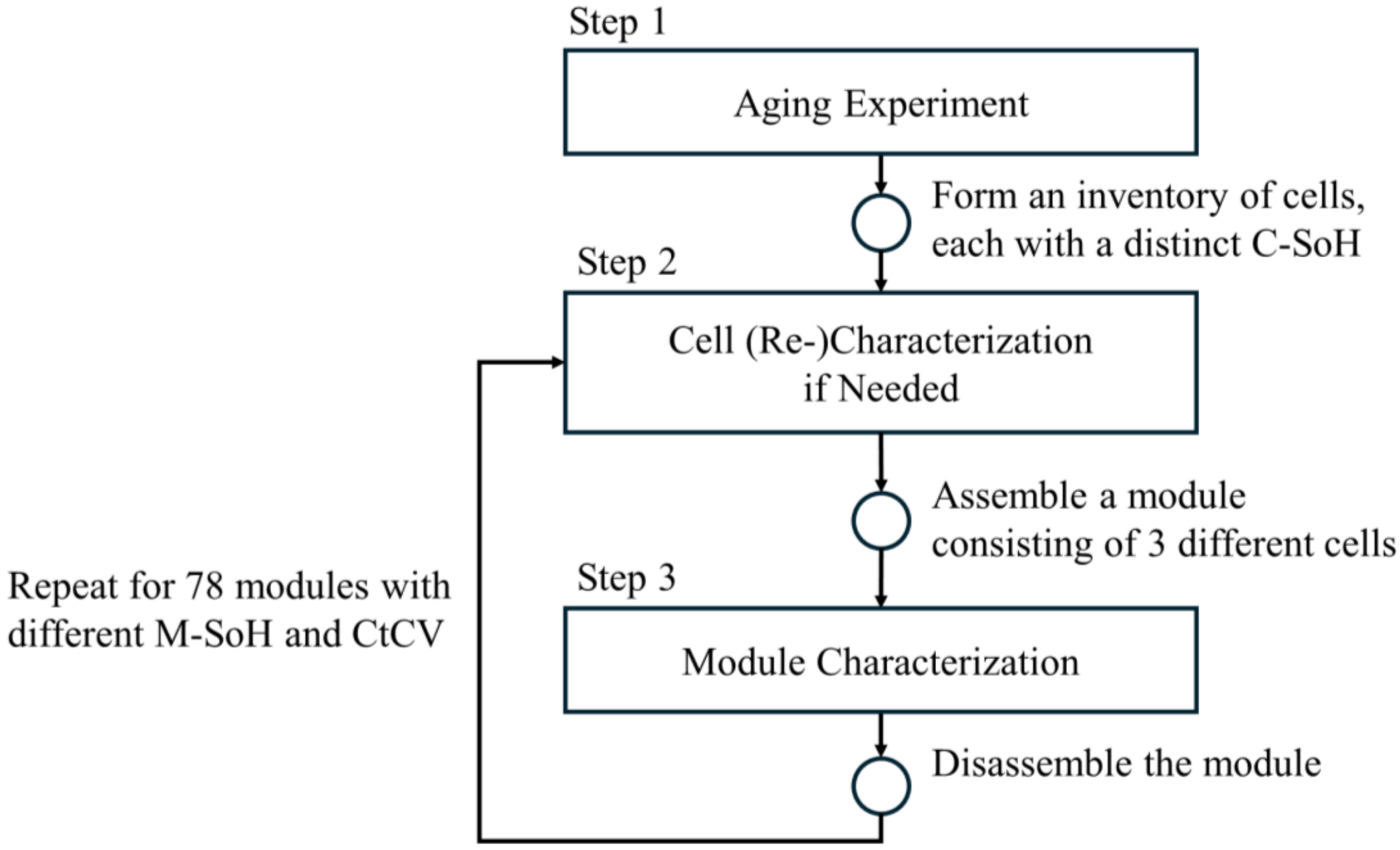

**Fig. 6. High-Level Flowchart of Experimental Procedures to Create the Dataset**

The section is organized as follows. Subsection A describes the aging cycles applied to all the fresh cells purchased from a manufacturer. Subsection B explains the initial cell characterization test after aging cycles and the repeated cell re-characterization tests as cell ages after several module characterization tests. Subsection C elaborates the module characterization tests.

## A. Aging Experiment

Seventy (70) fresh and cylindrical Sony US18650VTC6 cells are purchased. To establish a pool of cells with diverse C-SoH values ranging approximately from 100% to 80%, 54 of the 70 newly purchased cells are randomly selected and subjected to multiple rounds of aging cycles, while the remaining 16 cells are kept as fresh cells for later use. This subsection deals with Step 1 in Fig. 6. Fig. 7 shows the flowchart of the aging procedures, while Fig. 8 illustrates a representative aging sequence for a typical cell.

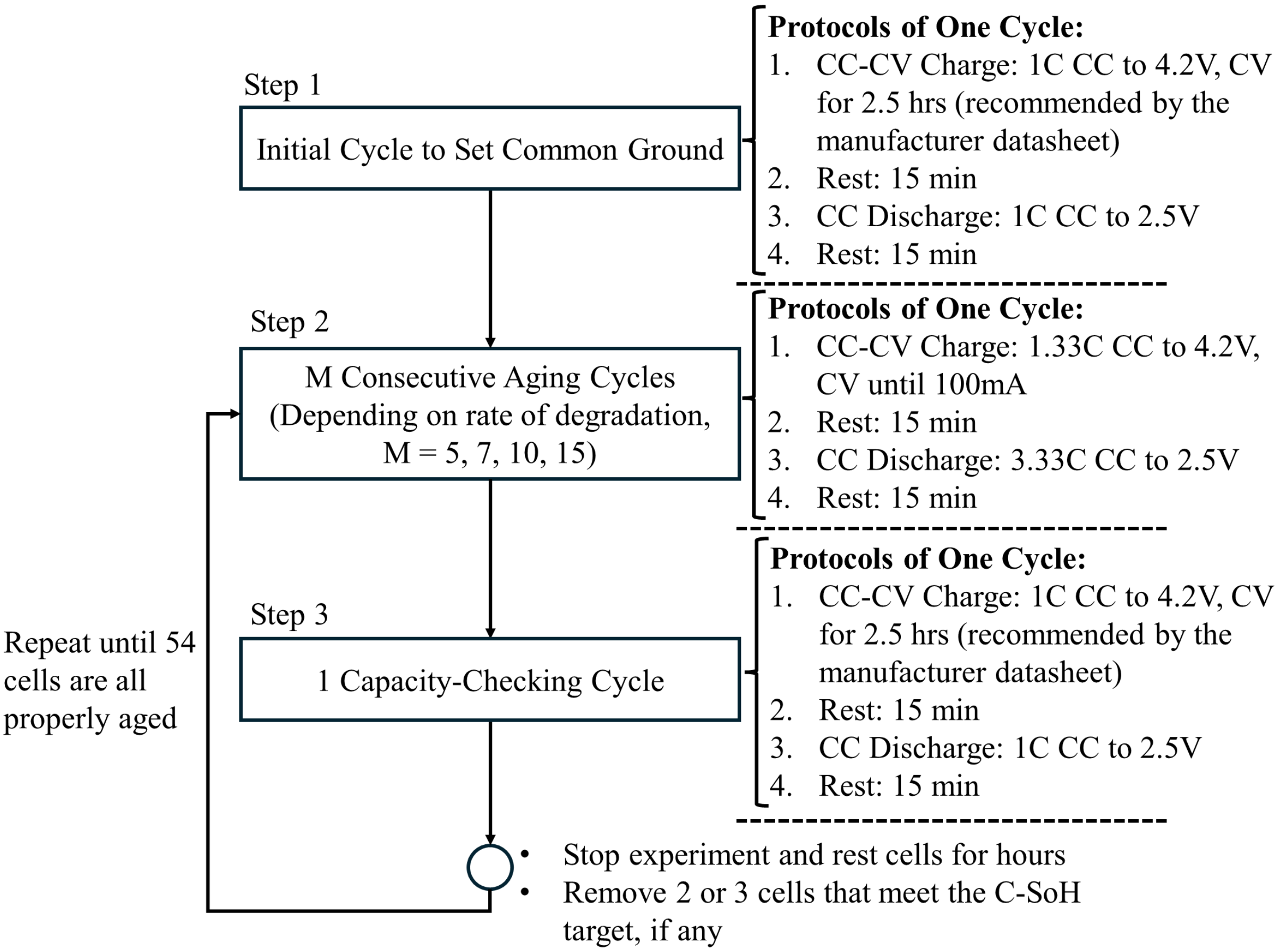

**Fig. 7. Flowchart of the Aging Procedures in the Dataset**

**Fig. 8. Example Aging Sequence for a Typical Battery Cell**

Based on Fig. 7 and Fig. 8, the aging procedure consists of three types of cycles: an initial cycle (step 1), an aging cycle (step 2), and a capacity-checking cycle (step 3). First, all 54 cells undergo an initial cycle to establish a common baseline. Since cells are purchased with different initial state of charge (SoC) and other states, this cycle aligns their conditions before the aging cycles. Second, all 54 cells are subjected to repeated rounds of several consecutive aging cycles. To accelerate degradation, every aging cycle consists of a constant-current-constant-voltage (CC-CV) charging at 1.33C and constant-current (CC) discharging at 3.33C. The number of consecutive aging cycles is progressively increased from 5 to 7, 10, and 15, as cell degradation slows at lower C-SoH levels. After each round of consecutive aging cycles, a cycle is performed to assess cell capacity, following the manufacturer-recommended protocol [4]. More accurate characterization is conducted later in Subsection B. Based on the capacity estimates, 2 or 3 cells are removed if they meet a target C-SoH, and the experiment resumes. The target C-SoH ranges approximately from 99% to 80% with 1% increment. Note that rest periods are intentionally kept short for the aging cycles to expedite the process. With these procedures, an inventory of 70 cells with C-SoH ranging roughly from 100% to 80% is obtained.

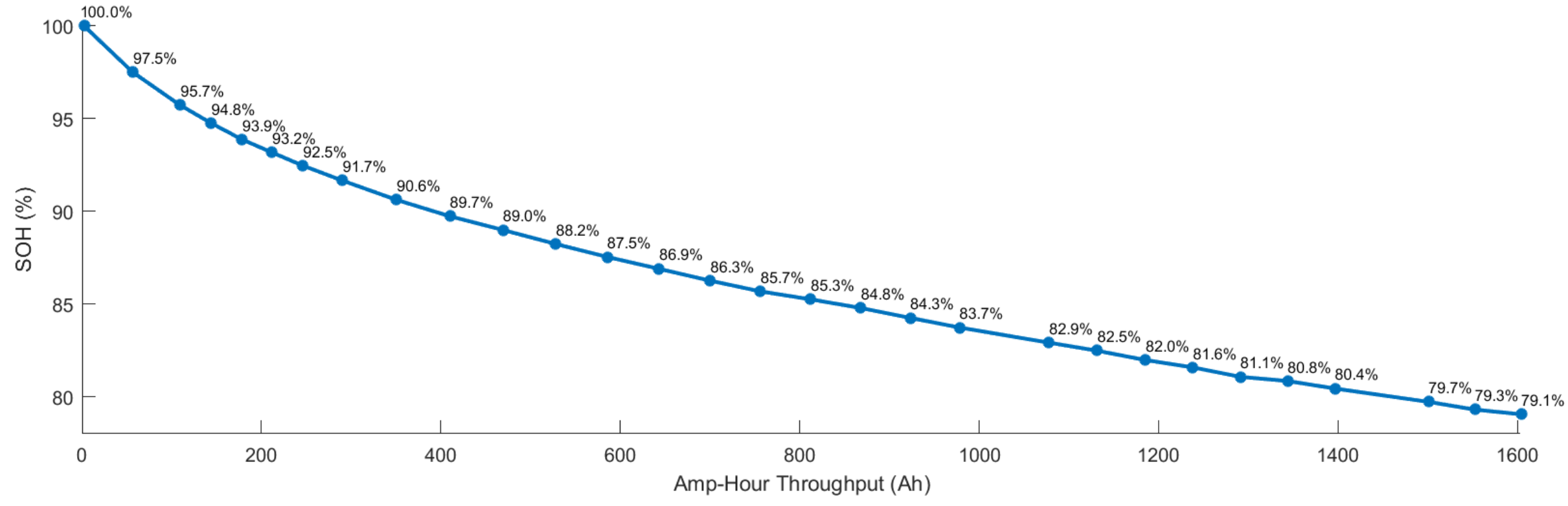

**Fig. 9. Cell Degradation Trend with Respect to Amp-Hour Throughput**

Fig. 9 shows a representative cell degradation trajectory as a function of cumulative ampere-hour (Ah) throughput. As illustrated in Fig. 9, the cell degradation rate decreases as C-SoH declines. This observation serves as a basis for Subsection B.2 of the *Experimental Design, Materials and Methods* Section.

## B. Initial Cell Characterization and Subsequent Cell Re-Characterization

This subsection corresponds to Step 2 in Fig. 6. It details the procedures of the initial cell characterization after aging and subsequent re-characterization and highlights the limitations of the cell re-characterization experiments.

### B.1. Initial Cell Characterization after Aging

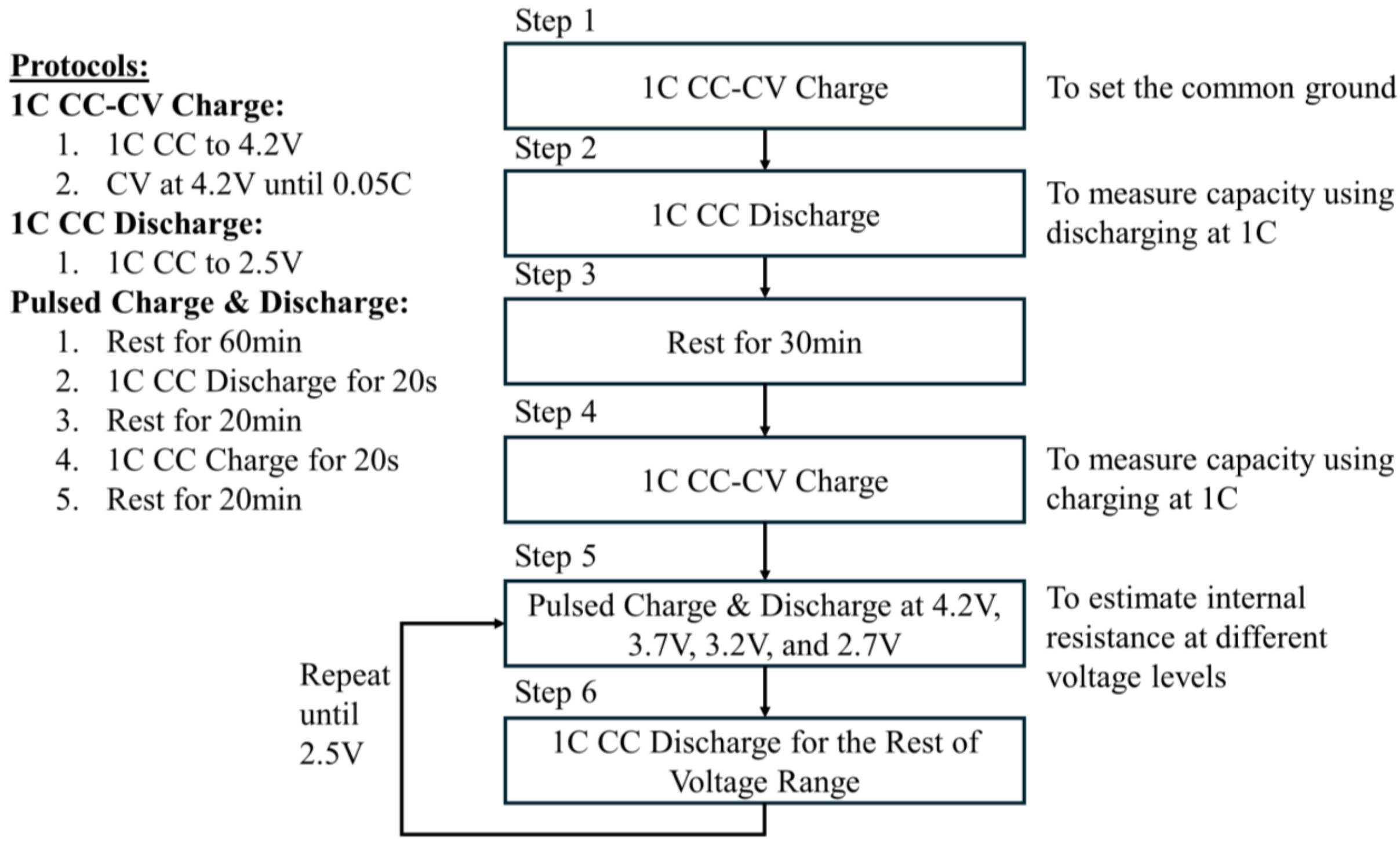

**Fig. 10. Flowchart of the Cell Characterization Procedures in the Dataset**

After the aging experiment, all 70 cells undergo an initial set of characterization cycles to quantify their capacities and internal resistances, prior to being assembled into modules. Fig. 10 illustrates the flowchart of the cell characterization procedures, while Fig. 4 in *Data Description* Section presents typical profiles measured during these tests. As shown in Fig. 10, the cell characterization consists of six steps. The first step standardizes the initial conditions, establishing a common baseline across cells with different initial states. The second and fourth steps correspond to 1C CC discharging and 1C CC-CV charging characterization tests, respectively. These tests enable the extraction of key metrics, such as C-SoH. A resting period (Step 3) is inserted between Steps 2 and 4 to reduce transient effects and allow the cell to relax. Steps 5 and 6 provide approximate estimates of cell-level internal resistance at several voltage levels, including 4.2V, 3.7V, 3.2V, and 2.7V. One limitation of Steps 5 and 6 is that, due to practical constraints, they are simplified from a standard hybrid pulse power characterization (HPPC) test, which typically estimates internal resistance at different states of charge (SoC) rather than at fixed voltage levels.

### B.2. Subsequent Cell Re-Characterization after Module Characterization Tests

As the cells are subsequently reconfigured into different modules and subjected to various module characterization tests (i.e., Step 3 in Fig. 6), they continue to age over time. To account for this progressive degradation, periodic re-characterization cycles are performed. These re-characterization tests follow the same procedures as illustrated in Fig. 10 and described in Subsection B.1, while Fig. 4 in the Data Description section presents representative profiles measured during these tests.

Recognizing that battery degradation occurs gradually, re-characterization cycles are scheduled based on a predefined empirical plan instead of after every module characterization test, as summarized in Table 7. This schedule reflects a practical trade-off between experimental efficiency and the accuracy of cell-level labels. As shown in Fig. 9, the degradation rate decreases as C-SoH declines. Consequently, from an efficiency perspective, to maintain a comparable level of accuracy for CtCV labels, cells with lower C-SoH require less frequent re-characterization than cells with higher C-SoH. With trial and error, we observed that, under the schedule, cells degrade by approximately 1% in SoH between two consecutive re-characterizations. Nonetheless, this re-characterization schedule may introduce an inherent limitation to the dataset. Because cell-level properties are only updated periodically, labels for CtCV and C-SoH may contain inaccuracies due to unobserved degradation between successive characterization points.

**Table 7. Empirical Schedule for Cell Re-Characterization**

| Condition | Frequency of Cell Re-Characterization |
|---|---|
| C-SoH of the target cell $\in [94\%, 100\%]$ | Every 1 module characterization test |
| C-SoH of the target cell $\in [88\%, 94\%)$ | Every 3 module characterization tests |
| C-SoH of the target cell $< 88\%$ | Every 6 module characterization tests |

## C. Module Characterization Tests

After initial cell characterization and repeated re-characterization tests, 78 modules with different M-SoH and CtCV values are assembled. This section corresponds to Step 3 in Fig. 6. Fig. 11 illustrates the flowchart of the module characterization procedures, while Fig. 3 in *Data Description* Section presents typical profiles measured during these tests. As shown in Fig. 11, each module undergoes three cycles, all of them under 25°C ambient temperature. The first cycle standardizes initial conditions, ensuring a common baseline across modules with different initial states. The second and third cycles correspond to characterization tests at 0.5C and 0.25C, respectively. These two cycles enable extraction of key information, such as M-SoH, and support diagnostic analyses, including incremental capacity analysis and differential voltage analysis.

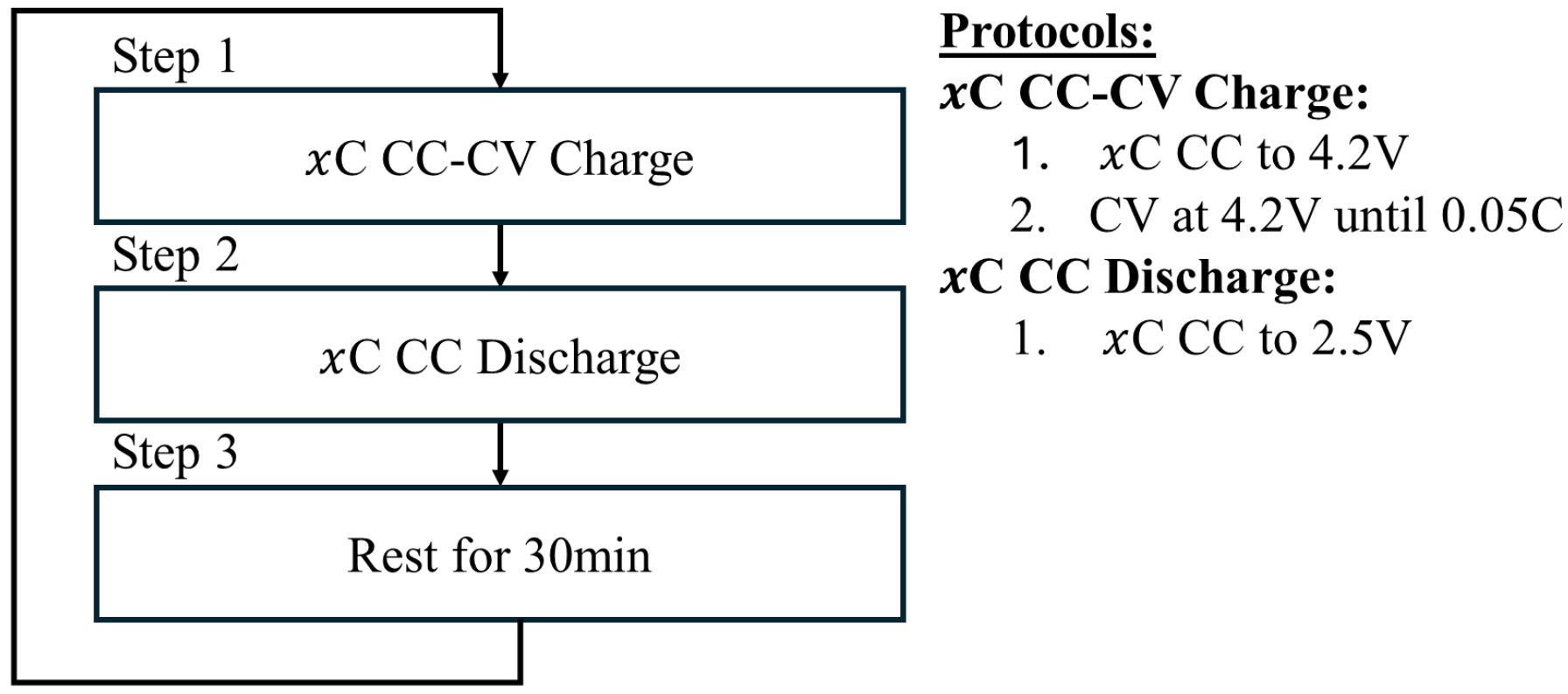

Repeat the loop 3 times:

- 1st Time: $x = 0.5$, to set the common ground
- 2nd Time: $x = 0.5$, to obtain characterization data under 0.5C
- 3rd Time: $x = 0.25$, to obtain characterization data under 0.25C

**Fig. 11. Flowchart of the Module Characterization Procedures in the Dataset**

## D. Experimental Apparatus

The setup of experimental apparatus is shown in Fig. 12. The battery cells used are Sony US18650VTC6 cells, with specifications provided in [4], as discussed earlier. All aging, cell characterization, and module characterization tests are conducted using a Maccor Series 4000 (S4000) battery cycler, with specifications given in [6]. Signals are measured at a sampling rate of 1Hz by Maccor S4000. Measured data are transferred via Ethernet connections to and stored in a computer.

For the cell aging and (re-)characterization tests, the cells are first assembled using standard molded Kelvin-contact cell holders provided as part of Maccor's standard accessories, with detailed information given in [7], and then inserted onto the Maccor cycler, as shown in Fig. 12.

For the module characterization tests, the modules are first formed by assembling three cells into four-18650-in-parallel battery holders from the BeiLaMoo 18650-BH Series, with detailed information provided in [8]. The modules are then connected to the Maccor S4000 using clamps and cables, as shown in Fig. 12. To maximize contact area and minimize contact resistance, the fourth slot of each module holder is intentionally left empty, allowing its metal contacts (circled in Fig. 12) to be directly clamped. The interconnect resistances between two branches within a battery module holder is measured using a Hioki BT3563-01 multimeter [8] and is found to be 15 mΩ.

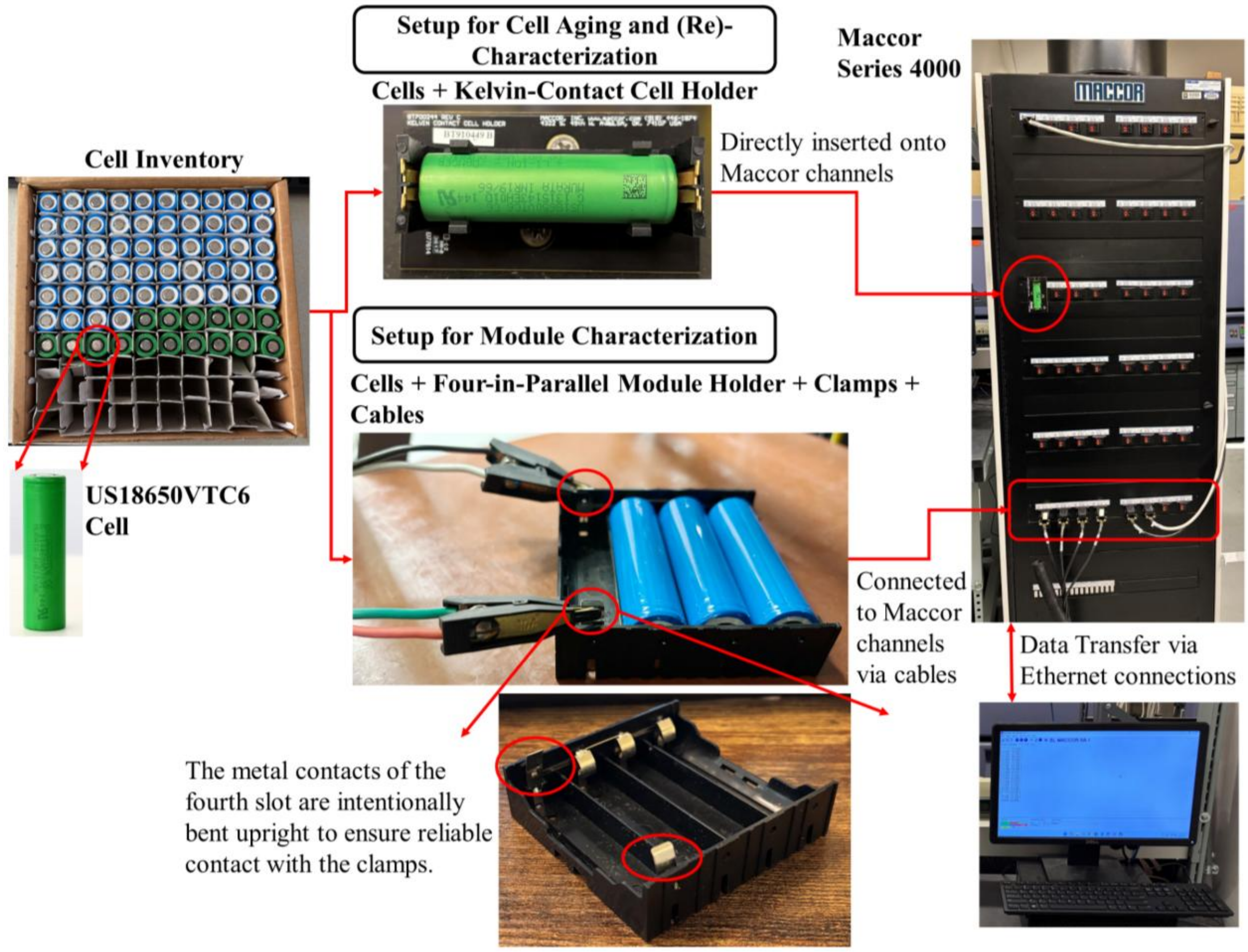

**Fig. 12. Setup of Experimental Apparatus**

## E. Summary

This dataset provides comprehensive experimental data for battery modules composed of three parallel-connected cells, spanning a wide range of module-level state of health (80.98% - 100% M-SoH) and cell-to-cell variations (0% - 9.31% SD). Crucially, it includes individual characterization data for every constituent cell within the modules. By pairing module-level measurements with precise cell-level labels, this dataset supports the development and validation of advanced diagnostic and prognostic algorithms for monitoring battery module degradation. However, several limitations must be acknowledged. First, because cell re-characterization was performed periodically according to a fixed schedule rather than before every module test, cell-level labels may contain errors due to unobserved degradation between intervals. Second, cell internal resistance was quantified via a simplified pulse test rather than the standard Hybrid Pulse Power Characterization (HPPC) method, potentially affecting resistance accuracy. Finally, the current dataset lacks complex real-world operating profiles, such as fast-charging protocols, dynamic loads, and standard driving cycles.

## LIMITATIONS

Although this dataset uniquely provides characterization data for both battery modules and their constituent cells across a wide range of M-SoH and CtCV values, it still has several limitations as detailed in the *Experimental Design, Materials, and Methods* Section. First, cell re-characterization is conducted only periodically according to a predefined schedule due to budget constraints, which inevitably introduces errors in the cell-level labels within each module. Second, due to practical constraints, the internal resistance of individual cells is measured using a simplified pulse test rather than a standard HPPC test. Third, more complex charging and discharging profiles encountered in real-world battery-powered systems, such as fast charging and urban driving cycles, are not included.

## ETHICS STATEMENT

The authors have read and follow the ethical requirements for publication in Data in Brief. The authors confirm that the current work does not involve human subjects, animal experiments, or any data collected from social media platforms.

## CRediT AUTHOR STATEMENT

**Qinan Zhou**: Conceptualization, Methodology, Software, Validation, Formal Analysis, Investigation, Data Curation, Writing – Original Draft, Visualization. **Daniel Stephens**: Conceptualization, Methodology, Investigation, Writing – Review & Editing. **Jing Sun**: Conceptualization, Resources, Writing – Review & Editing, Supervision, Project administration, Funding acquisition.

## ACKNOWLEDGEMENTS

The authors would like to sincerely thank Dr. Gregory Less and Andini Jinggan Ayumurti of the University of Michigan for their valuable support in experiments and contributions to this work.

This research did not receive any specific grant from funding agencies in the public, commercial, or not-for-profit sectors.

## DECLARATION OF COMPETING INTERESTS

The authors declare that they have no known competing financial interests or personal relationships that could have appeared to influence the work reported in this paper.